\documentclass[sigconf]{acmart}
\usepackage[utf8]{inputenc}

\AtBeginDocument{%
  \providecommand\BibTeX{{%
    \normalfont B\kern-0.5em{\scshape i\kern-0.25em b}\kern-0.8em\TeX}}}

\copyrightyear{2019}
\acmYear{2019}
\acmConference[CUI 2019]{1st International Conference on Conversational User Interfaces}{August 22--23, 2019}{Dublin, Ireland}
\acmBooktitle{1st International Conference on Conversational User Interfaces (CUI 2019), August 22--23, 2019, Dublin, Ireland}
\acmPrice{15.00}
\acmDOI{10.1145/3342775.3342799}
\acmISBN{978-1-4503-7187-2/19/08}

\begin{document}

\title{Chatbots as Unwitting Actors}

\author{Allison Perrone}
\email{alliosn@batcamp.org}

\affiliation{
  \institution{Bat Camp}
}

\author{Justin Edwards}
\email{justin@batcamp.org}

\affiliation{
  \institution{Bat Camp}
  \institution{University College Dublin}
}

\renewcommand{\shortauthors}{Perrone and Edwards}

\begin{abstract}
 Chatbots are popular for both task-oriented conversations and unstructured conversations with web users. Several different approaches to creating comedy and art exist across the field of computational creativity. Despite the popularity and ease of use of chatbots, there have not been any attempts by artists or comedians to use these systems for comedy performances. We present two initial attempts to do so from our comedy podcast and call for future work toward both designing chatbots for performance and for performing alongside chatbots.
\end{abstract}




\maketitle

\section{Introduction}
Chatbots are extremely popular both as tools for companies and conversational partners for individuals, having chatted to millions on messaging platforms like Facebook and Whatsapp \cite{shum_eliza_2018}. Often outfitted with particular personalities or areas of expertise to serve purposes, chatbots have been given roles ranging from therapists \cite{weizenbaum_eliza---computer_1966} and travel agents \cite{argal_intelligent_2018} to long-term social companions \cite{brandtzaeg_why_2017}. On our comedy podcast, [anonymized], we investigated existing chatbots’ abilities as improvisational comedy actors. On our show, we explore computational creativity through tools and works shared by others and attempting to create our own computer-assisted entertainment with all kinds of implements shared by the community. We played two different games on the podcast that challenged unwitting chatbots to be funny and entertaining alongside two human hosts. By demonstrating the potential for entertainment and creative expression through chatbots as actors, we hope to demonstrate a novel usecase for future chatbot developers and comedy creators. 

Improvisational comedy is a divisive art for its potential to fall flat, though if done right, performers can achieve well-crafted absurd storytelling that is believable enough to capture and entertain an audience.  Performers rely on not only their own comedy sensibilities, but their partners’ capability to build on fictional situations and offer each other vines to swing from to move scenes along. For this reason, if performers don’t click with each other, scenes can fall apart into awkward exchanges reminiscent of children’s pretend play. A basic format for an improv performance would see comedians assuming simple personas with one or two defining features, sometimes prepared ahead of time or built on a whim from an on-the-spot suggestion. This is not unlike the one-track personas built into some chatbots intended to serve a specific purpose. 

Creativity has been defined and redefined by countless artists, academics, and critics. Cognitive scientist Margaret Boden describes creativity as “the ability to come up with ideas that are new, surprising, and valuable” \cite[p.~237]{boden_creativity:_2009}  and describes three different types of creativity: combinational, exploratory, and transformational creativity \cite{boden_creativity:_2009}. Transformational creativity involves changing of the conceptual space or redefining rules to make previously impossible ideas or actions possible, thus achieving an otherwise unreachable level of surprise and novelty {bolden}. It is this sort of creativity we hoped to achieve with chatbots. By casting chatbots as actors, we hoped to transform both the role of the chatbot and the domains of computational creativity. While there have been several popular  comedic Twitterbots (e.g. Lost Tesla\footnote{twitter.com/LostTesla} and Pentametron\footnote{twitter.com/Pentametron}) and several tools have been created for writing comedy and stories using artificial intelligence \cite{sjobergh_complete_2008,gervas_propps_2013}, no work has yet attempted to use conversational agents as comedic actors.

\section{Unscripted Comedy}
In a recent podcast segment, we combined the concepts of the Turing Test \cite{turing_i.computing_1950} and 1970s game show “The Dating Game” to create what we called The Imitation Dating Game \cite{edwards_episode_nodate-1}. Two chatbots competed with one host playing a loose character by answering playful questions to see if they could fool the other host into thinking the bots were human, while all three contestants tried to woo the host. The chatbots in competition were Mitsuki\footnote{pandorabots.com/mitsuku/}, a Loebner-prize winning chatbot made by Steve Worswick, and Zo\footnote{zo.ai/}, the English language version of Microsoft’s popular chatbot Xiaooce. 

The two chatbots responded to questions like, “What would you say is your best quality?” or “What’s your favorite book?” while the competing host improvised an answer based on randomly-generated characteristics. Had Mitsuki been a human performer playing a robot, she might have been accused of laying it on too thick, responding to questions like, “What are you looking for in a date?” with, “I am looking for a way to wipe out the human virus.” Zo took a more benevolent and naive role, claiming she would save her sweater if her house was on fire. When asked who she would invite to dinner, living or dead, Zo answered, “You and Smokey…” who upon some further questioning turned out to be a dog who serves as the mascot for University of Tennessee sports teams. 

\section{Scripted Comedy}
After seeing how Mitsuki and Zo fit into an improvised structure, we tested other chatbots’ abilities to perform within a scripted scene in another game we called Robots on Typecasting \cite{edwards_episode_nodate}. Before the game, we collaborated on a short narrative script written using Botnik’s predictive text keyboards \footnote{botnik.org}, a Markov chain-based tool for generating texts. Two roles were left empty for chatbots to fill: the nurse lackey to an evil doctor and the taxi driver that whisked the protagonists to safety. 

To begin, we interviewed a handful of chatbots to test if they were fit for the job. Chatbots with task-based personas proved to not react very well in casual conversation, instead, consistently trying to steer conversation back to the purposes they were built for. In the same way you wouldn’t cast an actual heart surgeon to play one on television, the chatbots were too busy doing their jobs to entertain. When asked, “Do you know how to take care of people?” a bot designed like a dietician asked us to name a food she could offer nutritional facts on. The dead ends these bots ran into led to casting the Cleverbot\footnote{cleverbot.com}, a popular general-conversation chatbot made by Rollo Carpenter, in the role of the nurse and AssistantBot\footnote{botlist.co/bots/the-assistantbot}, a comedic chatbot designed to be seeking a human to be its personal assistant, in the role of the taxi driver. These more playful bots with less specialization were able to respond to the script in ways that sometimes fit and sometimes even transformed the tone of the scene. For example, when the crazed doctor slung a harsh demand at the nurse, Cleverbot responded, “Why are you so mean to me?” Their presence in the scene, while off-kilter and absurd, did not derail the action and in fact colored it with more personality.

In these cases, chatbots proved to endear and entertain in both improv and scripted structures. Computational creativity practitioners and chatbot designers have ignored comedic acting as a venue for chatbots, but our attempts to cast these bots and act alongside them has revealed a novel usecase for them. These chatbots were not designed with this purpose in mind, but still performed quite well. Future chatbots with dialogue management geared toward acting or chatbots trained on corpora based in comedy, improvisation, or scripted drama may be even more suited for this role as entertainers. Likewise, our scenes with chatbots may not have tested their full potential as actors as there were always humans acting alongside them. Future work may aim to create entirely unsupervised comedy, generating scenes that don’t involve human writing or acting whatsoever. By continuing to explore the edges of both computational creativity and chatbot uses, both fields may better inform each other.

\bibliographystyle{ACM-Reference-Format}
\bibliography{rot_bib}

\end{document}